\documentclass{elsart}
\usepackage[]{epsfig}
\begin{document}
 
\begin{frontmatter}
 
\title{Black holes production in self-complete quantum gravity}

\author{Euro Spallucci\thanksref{infn}}
\thanks[infn]{e-mail address: spallucci@ts.infn.it }
\address{Dipartimento di Fisica Teorica, Universit\`a di Trieste
and INFN, Sezione di Trieste, Italy}
 
\author{Anais Smailagic\thanksref{infn2}}
\thanks[infn2]{e-mail address: anais@ts.infn.it }
\address{INFN, Sezione di Trieste, Italy}
 
\begin{abstract}
A regular black hole  model, which has been proposed by Hayward
in \cite{Hayward:2005gi},  is reconsidered in the
framework of higher dimensional $TeV$ unification
and  self-complete quantum gravity scenario \cite{Dvali:2010bf,Dvali:2010ue,Spallucci:2011rn}.
We point out the ``quantum'' nature of  these objects and
compute their cross section production by taking into account
the key role played by the existence of a \textit{minimal length}
$l_0$. We show as the threshold energy is related to $l_0$.
We recover, in the high energy limit, the standard ``black-disk'' form of the cross section,
while it vanishes, below threshold, faster than any power of the invariant mass-energy 
$\sqrt{-s}$.
\end{abstract}
\end{frontmatter}

\section{Introduction}
 
Astrophysical size black holes are well modeled by classical
solutions of the Einstein equations.  Theoretical predictions
like the existence of  accretion disks and plasma jets along
rotation axis of spinning black holes have been confirmed by the amazing 
photos taken by orbiting telescopes.\\
Much different is the case of ``quantum black holes'' of microscopic size, 
whose description calls for a consistent theory of quantum gravity. 
String theory is presently the only framework providing a finite, anomaly-free,
perturbative formulation of gravity at the quantum level. However, even in 
this framework, black holes are usually treated as classical solutions
of certain Super-Gravity field theory emerging from a related 
Super-String  model in the ``point-like limit''. This approach is powerful 
enough to shed some light on the microscopic origin of black hole entropy 
(at least in certain cases) and to introduce higher dimensional objects, i.e.
$D$-branes, as solitonic solutions of the field equations. 
However, many answers are still to come, e.g. the form of the mass spectrum
of a quantum black hole, a satisfactory
resolution of the Information Paradox, an explanation of the horizon surface
quantization in elementary Planck cells, etc. All these problems 
are currently under an intensive investigation. Even more
compelling, is a satisfactory solution to the ``singularity problem'':
one expects that in a theory of extended objects, the very concepts
of (point-like) curvature singularity should be meaningless.
In other words, one would like to show that in a quantum theory of gravity
there should not be any curvature singularity neither ``naked'' nor
hidden by an event horizon \cite{Modesto:2004xx,Modesto:2007wp}. 
Unfortunately, this problem cannot be treated in any kind of
field theory (including Super-Gravity) where fundamental objects are
basically point-like.\\
At a first glance, this kind of physics could appear purely speculative and 
totally detached from any experimental verification. It may be true. 
However, $TeV$ quantum gravity is an intriguing
spin-off of non-perturbative String Theory, where all four interactions,
including gravity, are unified at an energy scale much lower than the Planck 
energy and, presumably, not too far away from LHC peak energy, i.e. $14\, TeV$.
The start of LHC runs opens the actual possibility to test
``new physics'' beyond the Standard Model, hopefully, including signatures
of ``quantum gravity'' phenomena 
\cite{Cavaglia:2002si,Rizzo:2006zb,Casanova:2005id,Casadio:2008qy,Gingrich:2010ed,Nicolini:2011nz,Bleicher:2010qr}.
\\
With this background in mind, we consider an effective approach to the 
singularity problem, where (semi)classical Einstein equations are used to 
determine black hole solutions ``keeping memory'' of their quantum nature.\\
A common feature of all candidate theory of quantum gravity is
the existence of a fundamental length scale where the very concept
of space-time as a classical manifold breaks down. Speaking of arbitrarily
small distances becomes meaningless and the concept of \textit{minimal length},
$l_0$, emerges as a new fundamental constant of Nature on the same footing
as the speed of light and Planck quantum of action \cite{Garay:1994en}.\\
String Theory \cite{Amati:1988tn,Fontanini:2005ik,Spallucci:2005bm}, 
non-commutative coordinates coherent states 
\cite{Smailagic:2003yb,Smailagic:2003rp,Smailagic:2004yy,Spallucci:2006zj,Nicolini:2010bj,Gaete:2011ka},
Generalized Uncertainty Principle \cite{Maggiore:1993rv,Konishi:1989wk}, 
Path Integral Duality \cite{Padmanabhan:1998yya,Shankaranarayanan:2000sp}, etc.,
share this common feature. \\
In reference \cite{Hayward:2005gi} an intriguing model of singularity-free black hole
was proposed in order to investigate, in a safe environment, back-reaction
effects of the Hawking radiation and the late stage of black hole evaporation.
There are several ways to change the form of the standard line
element in order to include quantum gravity effects 
\cite{Nicolini:2005vd,Ansoldi:2006vg,Nicolini:2008aj,Spallucci:2008ez,Nicolini:2009gw,Smailagic:2010nv,Nicolini:2011fy},
\cite{Modesto:2009ve,Hossenfelder:2009fc,Modesto:2010uh}
The  ``effective-model'' in \cite{Hayward:2005gi}  has several 
distinctive features. From our vantage point, we notice that
\begin{itemize}
\item it is mathematically simple and allows analytic calculations;
\item it encodes the basic features of more ``sophisticated'' models
of quantum gravity improved black holes, e.g. non-commutative geometry
inspired,  or Loop-quantum gravity black holes;
\item the mass spectrum is bounded from below by an extremal configuration;
\item the Hawking temperature vanishes as the extremal configuration is
approached.
\end{itemize}
The line element is given by

\begin{equation}
ds^2 = -\left(\, 1 -\frac{r_s\, r^2}{r^3+ r_s \,l^2_0}\,\right) dt^2 
+\left(\, 1 -\frac{r_s\, r^2}{r^3+ r_s \,l^2_0} \,\right)^{-1} dr^2 
+r^2\,\left(\,  d\theta^2 +\sin^2\theta d\phi^2 \,\right)\ ,
\label{hbh}
\end{equation}

where, $r_s= 2MG_N$ and $l_0$ is a new fundamental constant
on the same ground as $c$ and $\hbar$.  In order to keep
some degree of generality, we do not choose any specific value
for $l_0$ and consider it as a free, model-dependent, parameter
ranging from the $TeV$  up to the Planck scale.
\\
This metric is a solution of the Einstein equations with the following
energy-momentum tensor:

\begin{eqnarray}
&& T^0_0=-\rho=-\frac{1}{8\pi} 
\frac{3l^2_0 r_s^2 }{\left(\,r^3+ r_s \,l^2_0 \,\right)^2}\ ,
\label{rho}\\ 
&& T^r_r= p_r=-\rho\ ,\\
&& T^\theta_\theta=p_\theta=\frac{1}{8\pi} 
\frac{6l^2_0\, r_s^2\left(\,r^3- r_s \,l^2_0 \,\right)}
{\left(\,r^3+ r_s \,l^2_0 \,\right)^3}\ ,\\
&& T^\phi_\phi=p_\phi=p_\theta
\end{eqnarray}

The ``weak-point'' of the model is that the form of $T^\mu_\nu$ is not recovered
by an underlying theory, it is assumed in order to source the
(\ref{hbh}) field.\\
It is immediate to check that at short distance the metric is a  regular deSitter
geometry characterized by an \textit{effective} cosmological
constant of ``Planckian'' size $\Lambda_{eff}\equiv \frac{3}{l^2_0}$.
Thus, the curvature singularity is replaced by a core of ultra-dense
deSitter vacuum. \textit{This is the key
mechanism to stop matter to collapse into a singular, infinite density
configuration.} The deSitter vacuum plays the role of `` Planckian
foam'', a chaotic state of violent quantum gravitational fluctuations
disrupting the very fabric of the space-time continuum \footnote{
From a formal point of view, one can say that the ``regularization''
of the curvature singularity can be encoded into the substitution rule
\begin{equation}
 \frac{1}{r}\longrightarrow \frac{ r^2}{r^3+ r_s \,l^2_0}   
\nonumber
\end{equation}
 }.
In this paper we are going to provide a generalization of the
line element \ref{hbh} in the framework of $TeV$ quantum gravity
and compute the production cross section for this type of black holes,
by taking into account the existence of a minimal length. In this case,
the ``hoop-conjecture'' \cite{thorne} must be properly modified to comply with the
fact that an impact parameter smaller than $l_0$ is physically meaningless.

\section{$TeV$ ``quantum'' black hole}
 
In extending the metric (\ref{hbh}) in $d+1$ dimensions we choose a slightly different
definition of $l_0$ for reasons which will become clear later on

\begin{eqnarray}
&& ds^2 = - f\left(\, r\,\right)\, dt^2
+f\left(\, r\,\right) ^{-1} dr^2 
+r^2\,d\Omega_{d-1}\ ,\\
&& f\left(\, r\,\right)= 1 -\frac{2M G_\ast r^2}{r^d + 2 \frac{d-2}{d} M G_\ast l_0^2}
\label{extrabh}
\end{eqnarray}
where, $d\Omega_{d-1}$ is the infinitesimal solid angle in $d-1$ dimensions and
$G_\ast$ is the $TeV$-scale gravitational coupling with dimensions 
$\left[\, G_\ast\,\right]=L^{d-1}=E^{1-d}$. In our scenario the total mass
energy of the system equals the invariant mass of the two colliding partons,
i.e.

\begin{equation}
 M=\sqrt{-s}
\end{equation}

where $-s$ is the Mandelstam variable representing the center of mass energy squared.\\
The position of the horizon(s) is determined by the equation $f\left(\, r_H\,\right)=0$.
This is an algebraic equation of degree $d$. For arbitrary $d>3$ one can plot $M$ as a
function of $r_H$:

\begin{equation}
 M=\frac{1}{2G_\ast}\frac{r_H^d}{r_H^2-\frac{d-2}{d}\, l_0^2 }
\label{horizons}
\end{equation}

\begin{figure}[ht!]
\begin{center}
\includegraphics[width=10cm,angle=0]{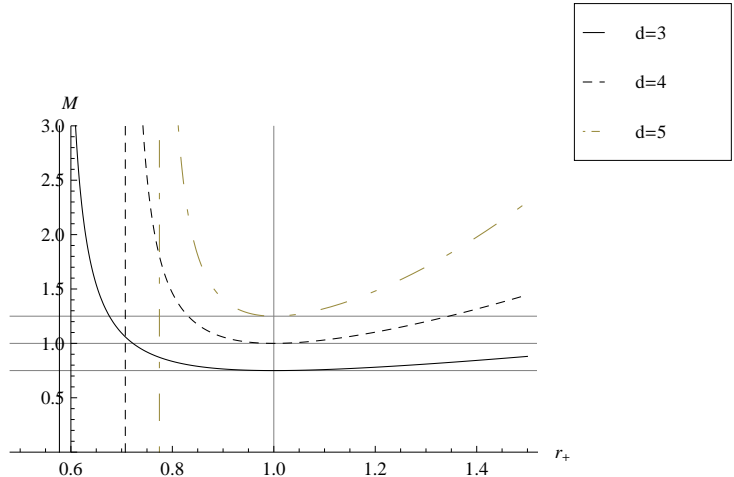}
\caption{Plot of $ M\left(\, r_H\,\right) $ vs $r_+$ for $d=3\ ,4\ ,5 $. The position of the minimum, which
correpsonds to the extremal black hole, is independent from $d$.  $l_0$ is our
unit of length. }\label{mass}
\end{center}
\end{figure}

For an assigned value of $M$ equation (\ref{horizons}) has two solutions, i.e. $r_H=r_\pm$,
 provided $M> M_0$, where $M_0$ is the absolute minimum of the function.
 The minimum corresponds to the
degenerate horizon of an extremal black hole and is determined by the two conditions
\begin{eqnarray}
 && f\left(\, r_0\,\right)=0\ ,\label{uno}\\
 && f^\prime \left(\, r_0\,\right)=0\ ,\label{due}
\end{eqnarray}

By solving the system (\ref{uno}),(\ref{due}) we gets
 
\begin{eqnarray}
 && r_0=l_0\ ,\label{qex}\\
 && M_0=\frac{1}{4G_\ast }\,\frac{d}{l_0^{d-2}}  \ ,\label{mex}
\end{eqnarray}

Equation (\ref{qex}) is a new and interesting result showing that 
\textit{there cannot exist black holes
with radius smaller than the minimal length.} This is not only a self-consistency check 
of our model, but
also the proof that extremal and near-extremal black holes are ``quantum'' object as 
their size falls in a  quantum gravity fluctuations dominated range. At the same time,
the existence of this kind of objects makes impossible to probe shorter distances.
It is important to recall that any further increase in energy makes black holes
bigger reducing the resolution power, and not viceversa as in the case of ordinary
particles.  This is the crux of the quantum gravity 
self-completeness scenario which has been recently discussed in 
\cite{Dvali:2010bf,Dvali:2010ue,Spallucci:2011rn}.  Furthermore, it is worth noticing that
$r_0$ is the same for any $d$  being determined by the  ``universal'' constant $l_0$ alone. \\
More in detail, we have three possible cases:
\begin{itemize}
 \item $M >M_0$ the metric has an (outer) Killing horizon, $r_H=r_+$, and an (inner) 
Cauchy horizon $r_H=r_-$ ;
 \item $M=M_0$ the two horizons coincide and the metric describe an \textit{extremal} black hole;
\item $M< M_0$ there are no horizon and the line element is sourced by a particle-like object.
If $M$ is not too small with respect to $M_0$, let us call these objects 
\textit{``quasi-black holes''} to
remark that they fall in the intermediate range of masses between particles and black holes.
 \end{itemize}
Thus, we meet a further nice feature of the model: it smoothly interpolates between
``particles'' and black holes by increasing the total energy of the system.
The transition between point-like objects and black holes is defined by the mass
$M_0$ of the extremal configuration. 
A detailed investigation of the thermodynamics properties of the black hole
(\ref{extrabh}) is postponed to a future paper, while we proceed in the next
section to the calculation of the production cross section.

\section{Cross section}
It is generally assumed that the production cross section for a black hole of
radius $r_H$ is simply its transverse area $\sigma\left(\, s\,\right)= \pi\,r_H^2\left(\, s\,\right) $.
Such a ``black disk'' cross section encodes the \textit{hoop-conjecture} \cite{thorne}: if two partons
collides with energy $\sqrt{-s}$ and \textit{impact parameter} $b$, black holes can be
produced if

\begin{equation}
 b\le r_H\left(\, s\,\right)
\end{equation}

Thus,  the scattering cross section for partons  of impact parameter $b$ reads

\begin{equation}
 \frac{d\sigma}{db}= 2\pi\, b\, \Theta_H\left[\, r_H\left(\, s\,\right)- b\,\right]
\end{equation}

where, $\Theta_H$ is the Heaviside step-function.  By integration over the un-observable
$b$ parameter, one gets the cross section

\begin{equation}
 \sigma\left(\, s\,\right)=2\pi\,\int_0^\infty db\, b\, \Theta_H\left[\, r_H\left(\, s\,\right)- b\,\right]
= \pi\,r_H^2\left(\, s\,\right)
\label{bdisk}
\end{equation}

As simple as that, this result is a little too naive. If equation (\ref{bdisk}) is literally taken,
we conclude that the probability to produce a black hole is non-zero even at arbitrary low  energy.
This is a result conflicting with all the known particle phenomenology. When a classical
argument like the hoop-conjectured is mismatched with a quantum cross section, the result
can often be unsatisfactory.\\
The root of the problem is that we let the impact parameter to range over arbitrary small values
while quantum gravity introduces sever limitations to the very concept of arbitrarily small
lengths.  In our case, it is meaningless to think about an impact parameter smaller than $l_0$.
We can translate this feature by requiring that our effective model of ``quantum'' black holes
breaks for $b< l_0$, and the cross section vanishes faster than any power of $s$. This kind of
asymptotic behavior can be achieved by introducing an exponential cut-off in the integration
measure in (\ref{bdisk})

\begin{equation}
 db\longrightarrow db\, e^{-l_0^2/ b^2}
\end{equation}

\begin{figure}[ht!]
\begin{center}
\includegraphics[width=8cm,angle=0]{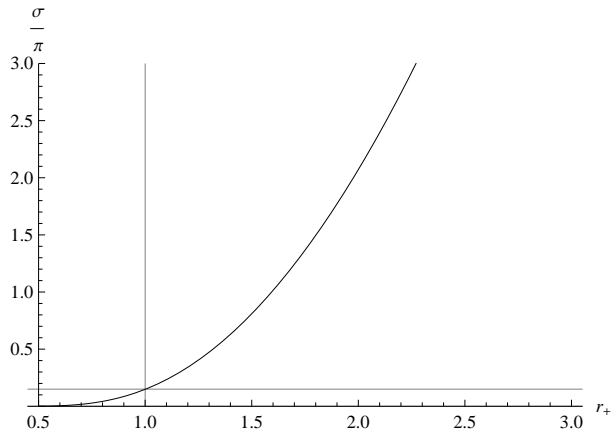}
\caption{Plot of $ \sigma\left(\, s\,\right)/\pi  $ vs $r_+$. We set $l_0=1$. 
The black disk limit is reached for $r_+>> 1$.}\label{asint}
\end{center}
\end{figure}

The resulting cross section reads

\begin{equation}
 \sigma\left(\, s\,\right)
= \pi\,l^2_0\, \Gamma\left[\, -1\ ; l^2_0/ r_H^2\left(\, s\,\right)\,\right]
\label{bdisk2}
\end{equation}

where, $\Gamma\left[\, -1\ ; l^2_0/ r_H^2\left(\, s\,\right)\,\right]$ is the upper incomplete
Gamma function which is defined as

\begin{equation}
 \Gamma\left(\, \alpha\ ; x\,\right)\equiv \int_x^\infty dt\, t^{\alpha -1}\, e^{-t}
\end{equation}

\begin{figure}[ht!]
\begin{center}
\includegraphics[width=8cm,angle=0]{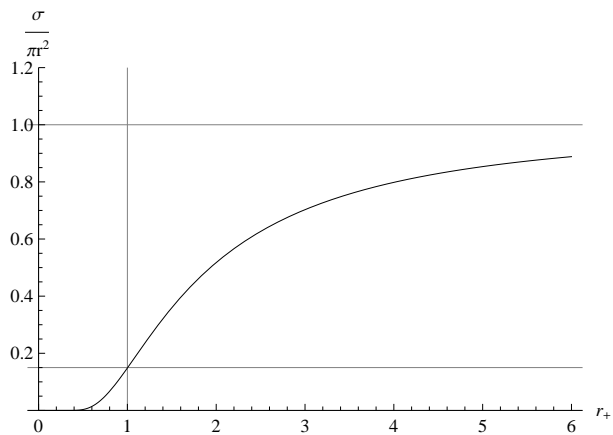}
\caption{Plot of $ \sigma\left(\, s\,\right)/\pi \,r_+^2 $. The horizontal asymptote represents
the black disk limit.}\label{asympt}
\end{center}
\end{figure}
At low energy, which is $\sqrt{-s}<< M_0 $, the cross section vanishes as

\begin{equation}
 \sigma\left(\, s\,\right)\approx 
\pi\,l^2_0\, \left(\,  \frac{r_H^4}{l_0^4}\,\right)\, e^{-l_0^2/r_H^2 }
\label{lowen}
\end{equation}

in agreement with our requirement. \\
In the opposite limit we need the following relation

\begin{equation}
 x^{-\alpha}\Gamma\left(\, \alpha\ ; x\,\right)\longrightarrow -\frac{1}{\alpha}\ , 
\quad x \longrightarrow 0
\end{equation}

Thus, for $\sqrt{-s}>> M_0 $, we recover the black disk cross section

\begin{equation}
 \sigma\left(\, s\,\right)\approx 
\pi\,l^2_0\, \left(\,  \frac{l_0^2}{r_H^2}\,\right)^{-1}=\pi\, r_H^2\left(\, s\,\right)
\label{highen}
\end{equation}

 The cross section {\ref{bdisk2}) smoothly interpolates between the (high energy) black disk limit
and an exponentially decreasing behavior below threshold. Immediately above threshold, it is energetically
preferred to produce  (near-) extremal black holes. The corresponding cross section
can be estimated to be 

\begin{equation}
 \sigma\left(\,\sqrt{-s}=M_0\,\right)=\pi\,l^2_0\,\Gamma\left[\, -1\ ; 1\,\right]\approx 0.15\,\pi\,l^2_0
\label{exbh}
\end{equation}

We have seen that below threshold black hole cannot exist. They are both
off-mass shell and too small to be considered as physical object. Thus,
we need to understand the physical meaning of the exponentially suppressed,
but non-zero, tail of the below threshold cross section. As our model smoothly
interpolates among different kind of physical objects 
\begin{center}
 particles
$ \longleftrightarrow $ quasi-black holes
$\longleftrightarrow $ black holes
\end{center}

we propose to continue  the cross section (\ref{lowen})  below threshold in the following way.
A quasi-black hole is a particle-like object characterized by a Compton wave-length $\lambda_C=1/\sqrt{-s}$.
From the vantage point of self-complete quantum gravity the transition between particle and black holes
implies to replace $\lambda_C(s)$ with $r_+(s)$ ( or, viceversa) as a characteristic
length of the object itself. In order to fit the asymptotic behavior (\ref{lowen}), 
 we replace  (\ref{bdisk2}) below threshold with

\begin{equation}
 \sigma\left(\, s\,\right)
= \pi\,l^2_0\, \Gamma\left[\, -1\ ; \lambda_C^2\left(\, s\,\right)/l_0^2\,\right]\ , \quad 
\label{belowth}
\end{equation}

Cross section (\ref{belowth}) smoothly joins to (\ref{bdisk2}) at the
critical point $\sqrt{-s}=M_0$ and exponentially vanishes at low energy as

\begin{equation}
 \sigma\left(\, s\,\right)
\approx \pi\,l^2_0\,\frac{l_0^4}{\lambda_C(s)^4}\, e^{-\lambda_C(s)^2/l_0^2}\ , \quad \lambda_C(s)>> l_0
\label{lowbelow}
\end{equation}

Thus, we can say that below threshold cross section (\ref{belowth}) describe the production
of quasi-black holes which are very massive objects with respect to ordinary particles, but not yet heavy enough
to collapse into black holes.  \\
It s worth to remark, that even after crossing the threshold  and approaching the geometric limit,
``Trans-Planckian'' scale is never probed as $r_+ > l_0$. The fate of near-extrenmal, non-thermal black holes
will be discussed in the next section.

\section{Conclusions and speculations} 

In this last section we would like to offer some speculations about one of the most intriguing
feature of black thermodynamics, i.e.  horizon surface quantization in elementary
Planck cells.\\
In spite of its simplicity, our model can offer a simple recipe for the mass spectrum
and the late stages of Hawking evaporation. The temperature

\begin{equation}
 T_H=\frac{d-2}{4\pi\, r_+}\left(\, 1-\frac{l_0^2}{r_+^2}\,\right)\ ,\quad r_+\ge l_0
\end{equation}

vanishes as  the genuine quantum regime is approached, i.e. $r_+\to l_0$. In this phase,
thermal behavior is negligible and we expect  different decay modes to provide the dominant contribution.\\ 
 The crux of our argument is once again
relation (\ref{qex}). As $l_0$ is the minimal length, the area of the event horizon
of the extremal configuration is entitled to be seen as the  fundamental ``\textit{quantum of area}''

\begin{equation}
 A_0=4\pi\, l_0^{d-1} \label{qarea}
\end{equation}

From this vantage point, we can see any non-extremal black hole as an ``\textit{areal excitation}''
 of the  \textit{ground state} given by the extremal configuration, in the sense that the 
{the area of the event horizon is an integer multiple of } $A_0$.  This is our quantization
condition:
\begin{equation}
A_H= 4\pi\, n\, l_0^{d-1}\ ,\qquad n=1\ ,2\ ,3\ ,\dots  \label{quantarea}
\end{equation}

The relation (\ref{quantarea}) takes into account the cellular structure of the horizon where
the quantum of area $ 4\pi\, l_0^{d-1} $ plays the role which is generally assigned to the Planck area
$l_{Pl.}^2 $.\\
From equation (\ref{quantarea}) it is immediate to derive the quantization rule for
the event horizon radius

\begin{equation}
r_+=  n^{1/(d-1)}\, l_0 \label{qradius}
\end{equation}
 
Finally, by inserting (\ref{qradius}) in (\ref{horizons}) one gets the mass spectrum

\begin{equation}
M_n = \frac{l_0^{d-2}}{2G_\ast}\,\frac{n^{d/(d-1)}}{n^{2/(d-1)}-(d-2)/d }\label{mspectr}
\end{equation}

The mass spectrum is bounded from below by the mass of the extremal black hole, i.e. 
$ M_{n=1}=M_0$, and approaches a continuum in the semi-classical limit $n >> 1$.\\
The emerging scenario suggests that large black holes decay thermally, while small objects decay
quantum mechanically by emitting quanta of energy $ \delta M= M_{n+1} -M_{n}$. Thus, the late stage
of the black hole evolution turns out to be quite different from the semi-classical thermal emission
and much more similar to the decay of an ordinary unstable particle. On a qualitative ground, this
conclusion is in agreement with the results obtained in \cite{Meade:2007sz}, where it is shown
that quantum black holes decay into a limited number of particles estimated to be between six and
twenty.

\end{document}